\documentclass[runningheads]{llncs}
\usepackage[T1]{fontenc}
\usepackage{graphicx}
\usepackage{tcolorbox}
\usepackage{relsize}
\usepackage{xspace}
\usepackage{subfiles}
\usepackage{booktabs}
\usepackage{subfiles}
\usepackage{graphicx}
\usepackage{multicol}
\usepackage{amsmath}
\usepackage{amsfonts}
\usepackage{booktabs,makecell,multirow,tabularx}
\usepackage{subfig}
\usepackage{csquotes}
\usepackage[hidelinks]{hyperref}
\usepackage{float}
\usepackage{placeins}

\begin{document}
\definecolor{mycolor}{rgb}{0,0,0}
\title{Variable Rate Image Compression via N-Gram Context based Swin-transformer}
%
%
\author{}
\authorrunning{}
\institute{}

\author{Priyanka Mudgal}
\authorrunning{Mudgal et al.}
%
\institute{Portland State University, Portland OR 97124, USA \\
\email{pmudgal@pdx.edu}}

%

\maketitle              
%

%
%
%
\begin{abstract}
This paper presents an N-gram context-based Swin Transformer for learned image compression. Our method achieves variable-rate compression with a single model. By incorporating N-gram context into the Swin Transformer, we overcome its limitation of neglecting larger regions during high-resolution image reconstruction due to its restricted receptive field. This enhancement expands the regions considered for pixel restoration, thereby improving the quality of high-resolution reconstructions. Our method increases context awareness across neighboring windows, leading to a -5.86\% improvement in BD-Rate over existing variable-rate learned image compression techniques. Additionally, our model improves the quality of regions of interest (ROI) in images, making it particularly beneficial for object-focused applications in fields such as manufacturing and industrial vision systems.

\end{abstract}

\section{Introduction}
\label{sec:intro}

\vspace{-0.2cm}
\begin{figure}
\centering
\vspace{-3.0cm}
\includegraphics[width=1\textwidth, trim = 0cm 8cm 13cm 0cm]{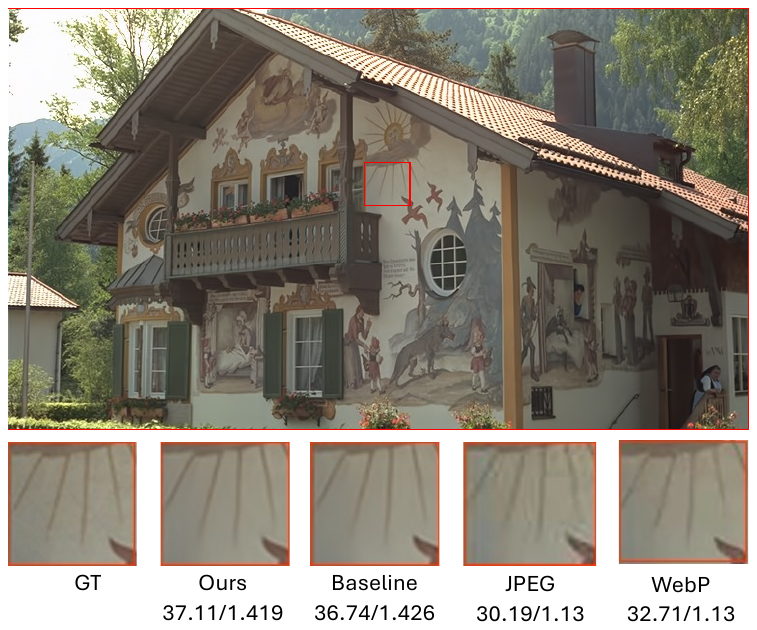}
\caption{\label{fig:baseline_comparison}The visualization of the kodim24 reconstruction from the Kodak dataset shows that our method achieves better PSNR while maintaining or reducing the bit-rate compared to baseline \cite{10222853} and traditional  methods. The subtitles indicate \textcolor{mycolor}{PSNR↑/bpp↓.}}
\vspace{-0.6cm}
\end{figure}

In recent years, learned image compression (LIC) methods have significantly advanced, surpassing traditional techniques in both efficiency and quality. Inspired by early research \cite{toderici2016variable,balle2016density}, modern LIC approaches, particularly those based on variational autoencoders (VAE) \cite{zou2022devildetailswindowbasedattention,mudgal2024enhancing}, optimize image compression by learning end-to-end representations tailored to minimize rate-distortion (RD) loss. However, most LIC models are optimized for fixed compression rates, requiring separate models for each bit-rate, which can limit their real-time application.

To address this issue, several variable-rate LIC techniques have been proposed to adjust bit-rates through additional parameters or algorithms \cite{song2021variableratedeepimagecompression,10222853,zhang2019learnedscalableimagecompression}. For instance, the spatially adaptive rate control in \cite{song2021variableratedeepimagecompression} and the vision transformer-based model in \cite{9770776} offer improvements in compression efficiency but encounter issues like time-consuming back-propagation. Other approaches in \cite{cui2022asymmetricgaineddeepimage,tong2023qvrfquantizationerrorawarevariablerate}, adjust quantization step sizes and gain factors to control bit-rate, yet still require training multiple models for effective rate control across various bit-rates. Kao et al. \cite{10222853} introduced a Swin Transformer-based model with Window-based Self-Attention (WSA) to combine long-range dependencies with the locality of convolutions. However, the small receptive field in WSA limits the model’s ability to capture fine details and textures, leading to distorted reconstructions, particularly in complex areas. More recent paper, particularly, Feng et al. \cite{Feng_2025_CVPR} proposed a linear attention mechanism using bi-receptance weighted key value (Bi-RWKV) blocks and spatial-channel context modeling, achieving substantial BD-rate reductions. In parallel, Zhang et al. \cite{Zhang_2025_CVPR} approached rate-distortion optimization as a multi-objective learning problem, yielding consistent gains. Additionally, Tu et al. \cite{tu2025multiscaleinvertibleneuralnetwork} developed a multi-scale invertible neural network (MS-INN) that enables wide-range bit-rate control using a single model.

While these methods advance the field, they also present notable limitations. Feng et al.’s approach \cite{Feng_2025_CVPR}, although efficient, relies on RWKV blocks originally designed for sequential modeling, which may limit spatial granularity. Zhang et al.’s work \cite{Zhang_2025_CVPR} primarily improves training dynamics but lacks mechanisms for spatial adaptivity or perceptual quality enhancement. Furthermore, its reliance on fixed training priors may reduce generalization across diverse content. Tu et al.’s MS-INN \cite{tu2025multiscaleinvertibleneuralnetwork}, though achieving strong rate-distortion performance, introduces higher computational complexity and offers limited flexibility for region-specific or fine-detail compression due to the constraints of invertible architectures. These gaps highlight the need for a method that combines spatial adaptivity, computational efficiency, and fine-detail preservation within a single variable-rate model.

\begin{figure}
\centering
\includegraphics[width=0.5\textwidth, trim = 1cm 13cm 1cm 2.0cm]{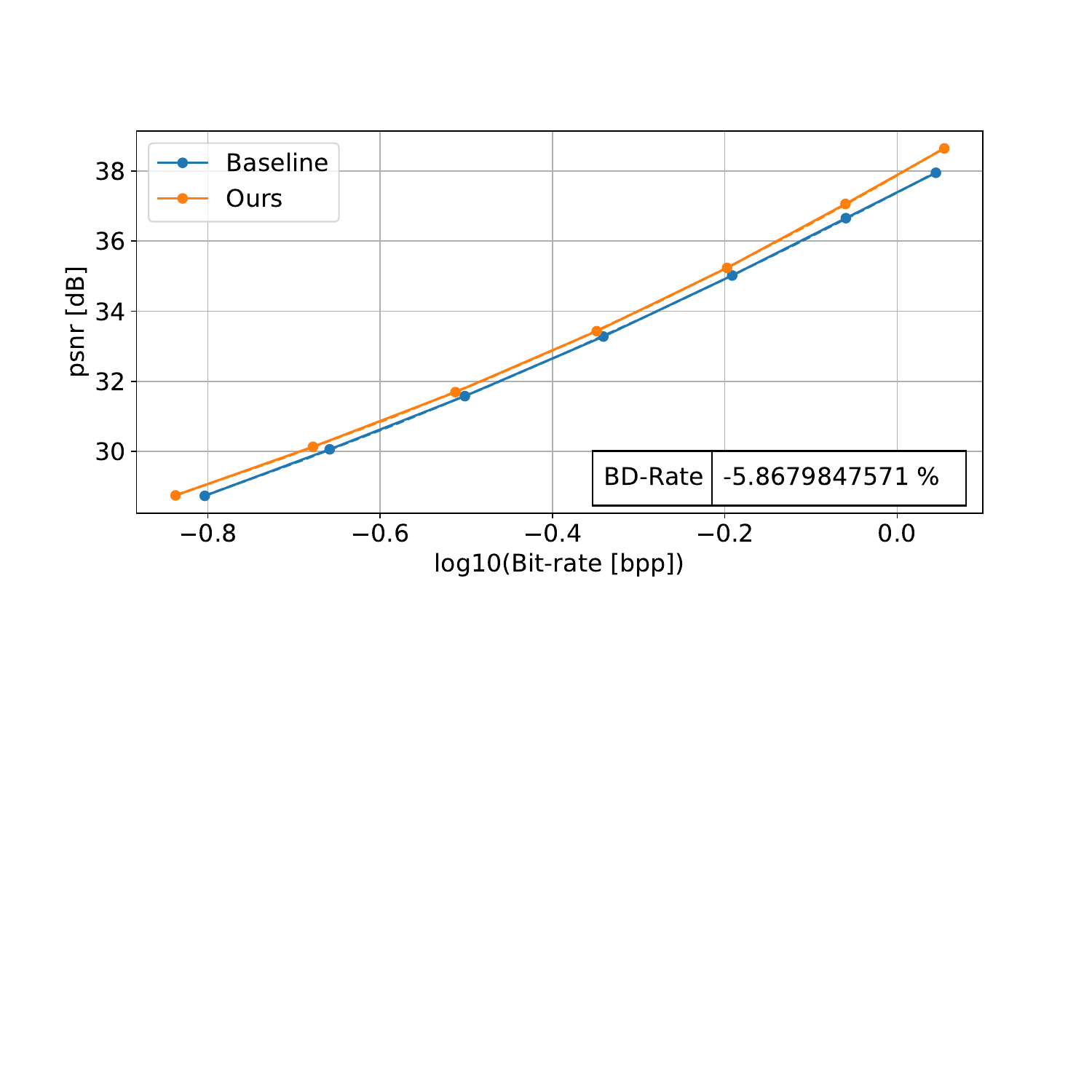}
\caption{\label{fig:bd-rate}BD-rate comparison of our proposed method using N-gram context with the baseline method \cite{10222853}.}
\end{figure}

In this work, we address these challenges by modifying the Swin Transformer block (STB) and incorporating N-gram context-based partitioning \cite{choi2023ngramswintransformersefficient} before applying WSA, enabling variable-rate compression using a single model. Inspired by the success of N-gram context in super-resolution \cite{choi2023ngramswintransformersefficient}, we extend this concept to image compression to better preserve high-frequency components and fine textures. This modification effectively expands the receptive field, enhancing the model’s ability to capture rich local and global context. Additionally, we apply sliding WSA to N-gram embeddings and reduce computational overhead using channel-reducing group convolutions. These improvements yield more accurate reconstructions and fewer compression artifacts, achieving a 5.86\% reduction in BD-rate, as shown in Fig. \ref{fig:bd-rate}. Furthermore, unlike prior works, we introduce an ROI-aware compression mechanism by selectively applying N-gram embeddings to semantically important regions offering spatial adaptability and perceptual control, which is not addressed in existing single-model variable-rate methods.

\vspace{-0.9cm}
\section{Proposed Method}
\label{sec:method}

\vspace{-0.2cm}

\begin{figure*}
\centering
\includegraphics[width=0.9\textwidth, trim = 0cm 5cm 0cm 4.2cm]{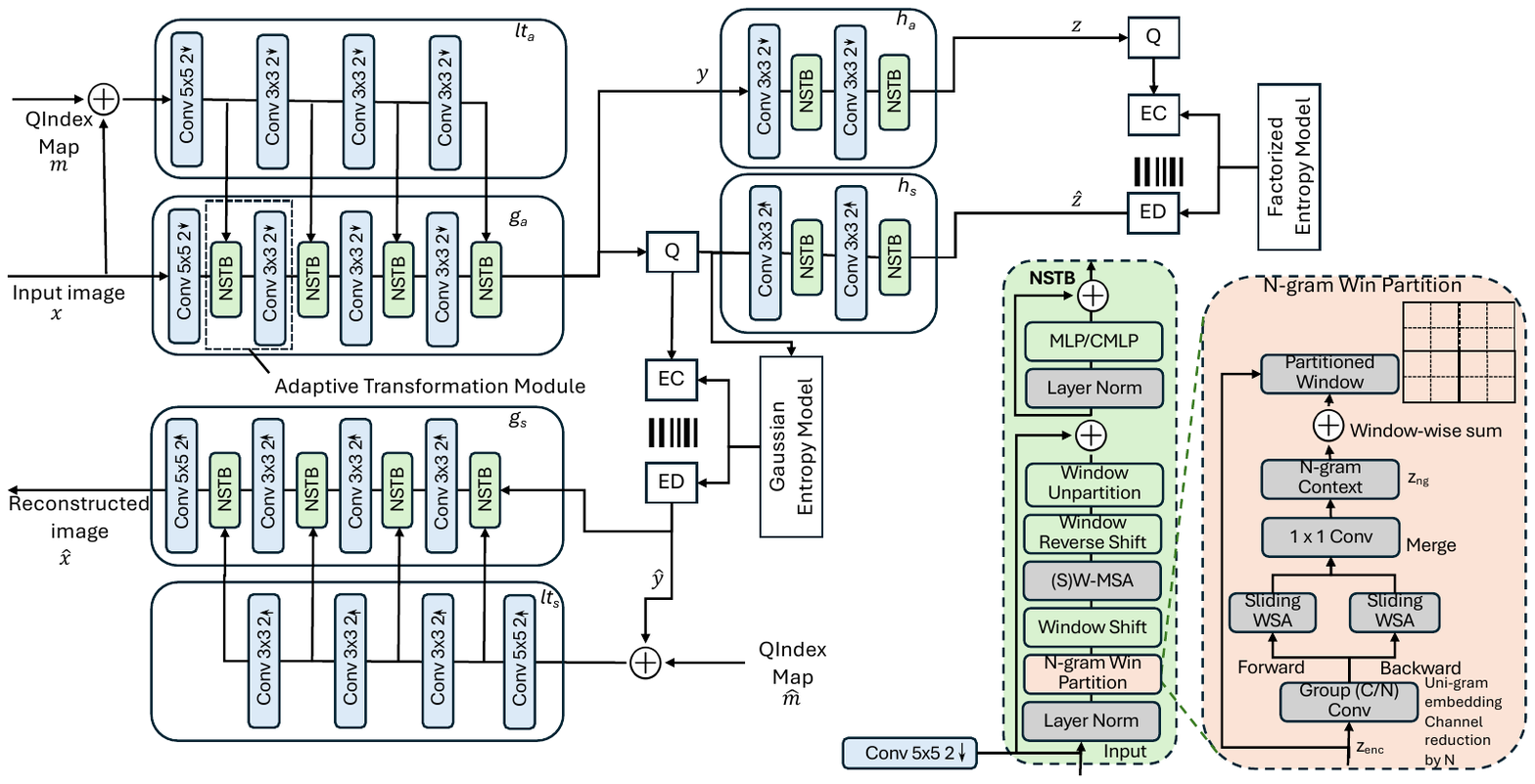}
\caption{\protect\label{fig:architecture} The architecture of \textcolor{mycolor}{our proposed} network is based on \cite{10222853}. The analysis $g_a$ and synthesis transform $g_s$ convert variables from image space (x) to latent space (y) and from latent space ($\hat{y}$) to image space ($\hat{x}$) respectively. EC and ED represent the arithmetic encoder and arithmetic decoder, respectively. The hyperprior analysis and synthesis transforms are from Minnen et al. \cite{ballé2018variational}. Blocks with dotted outline shows NSTB \textcolor{black}{adopted from \cite{choi2023ngramswintransformersefficient}}. It contains the uni-Gram embedding and sliding-WSA process. The dimensionality reduction via uni-Gram embedding enhances the efficiency of sliding-WSA. Bi-directional contexts share the same sliding-WSA weights. For window-wise summation, a value from $z_{ng}$ is added equally to $M^2$ pixels in a local window at the corresponding position.}

\end{figure*}

We propose an N-gram-based Swin Transformer image compression system that enables variable-rate compression with a single model and spatially adaptive quality control for regions of interest (ROI). The system architecture is shown in Fig. \ref{fig:architecture}. Our approach builds on the transformer-based image compression framework \cite{lu2021transformerbasedimagecompression,10222853}. The core autoencoder includes analysis $g_a$ and synthesis blocks $g_s$, as well as hyperpriors $h_a$ and $h_s$. Both encoding ($g_a$ and $h_a$) and decoding ($g_s$ and $h_s$) blocks feature N-gram Swin Transformer blocks (NSTB) interleaved with convolutional layers, as detailed in Section \ref{sec:nstb}.

\textcolor{black}{During encoding, the network receives the input image $x$ $\epsilon$ $R^{3\times H\times W}$ together with a QIndex map $m$.} For ROI-based compression, an ROI mask $r$ $\epsilon$ $R^{1\times H\times W}$ is also used to emphasize specific regions of the image. The QIndex map $m$ has values in the range $[0,1]$, dictating the bit-rate of the compressed latent representations. The ROI mask $r$, with values in $[0,1]$, acts as a weighting function to prioritize certain pixels for compression efficiency. \textcolor{black}{These inputs provide auxiliary information to the main encoder ${g_a}$, which produces the learned tokens.} Additionally, the QIndex map is input to $lt_a$, producing learned tokens that condition the NSTB and control the variable bit-rate. The image is first processed through a convolutional layer, then passed through a series of Adaptive Transformation Modules (ATMs). The hyper encoder $h_a$ follows the same structure but includes two ATMs. Each ATM consists of an NSTB followed by a convolutional layer, designed to capture both long-range and local dependencies in the image. These modules enable adaptive encoding, adjusting to varying levels of detail across the image, particularly for regions defined by the ROI mask $r$.

Before passing the input through the NSTB, a feature embedding layer projects the input features from size \( H \times W \times C \) to flattened dimensions of \( HW \times C \). In the NSTB, both image and learned tokens are processed together. The image tokens are augmented with learned tokens in the multi-head self-attention mechanism, where key and value matrices incorporate both types of tokens. This allows the attention mechanism to attend to both by concatenating them and applying attention across the windowed tokens. The resulting tokens are used for further processing. Then, N-gram context is applied before the shifted window attention mechanism. This block also includes a modified Multi-Layer Perceptron (MLP), using GELU activation with tanh approximation \cite{De_Ryck_2021}. We call this modified version Tanh-Approximate GELU MLP (TAG-MLP). The TAG-MLP layer computes window-based self-attention, and a feature unembedding layer remaps the attention-weighted features back to the original size of \( H \times W \times C \).

The synthesis module \( g_s \) handles the quantized image latent \( \hat{y} \) and a downscaled QIndex map \( \hat{m} \in \mathbb{R}^{1 \times \frac{H}{16} \times \frac{W}{16}} \) from \( lt_s \), matching the spatial resolution of \( \hat{y} \). It reverses analysis module's operations, restoring the original image features from the quantized representation, and predicting the latent's probability distribution more effectively and efficiently.

\subsection{\textbf{N-Gram Swin Transformer Block}}\label{sec:nstb}

We adopted NSTB from~\cite{choi2023ngramswintransformersefficient} and briefly describe it for the completeness of this paper. Please refer to \cite{choi2023ngramswintransformersefficient} for full details. \textcolor{black}{NSTB is based on scaled-cosine WSA, which operates within local windows of size \( M \times M \), where \( M \) is set to 8 by default. In this mechanism, the query, key, and value matrices represents the features of all pixels within a window. These matrices are used to compute pairwise similarities. Specifically, cosine similarity is calculated between each query and key pair to measure how closely they align. These similarity scores are then scaled by a learnable scalar \( \tau \), which controls the sharpness of the attention distribution and is initialized to values greater than 0.01, following the recommendation in~\cite{choi2023ngramswintransformersefficient}. A relative position bias matrix \( B \) is also added to incorporate spatial information. The resulting scores are normalized using the softmax function. The final attention output is obtained by applying these weights to the value matrix. This allows the model to capture rich dependencies among pixels within the local window.}




\textcolor{black}{In the window partitioning shown in Fig. \ref{fig:architecture}, we integrate the N-gram context algorithm adopted from~\cite{choi2023ngramswintransformersefficient} and following their approach to plug it in our network. First, we map the input image to uni-gram representation, which reduces the number of channels and image resolution. Then, we compute the forward N-gram feature by setting \( M = N \) and \( D = D/2 \). During this step, the sliding-WSA is implemented as a sliding-window convolution followed by N x N average pooling. Next, the forward and backward N-gram features are concatenated, after which a \( 1 \times 1 \) convolution combines them to generate the N-gram context. Then, the N-gram context, \( z_{ng} \), is added to each window of the image, with the same value applied to all pixels within a window. This adjusts the pixels based on average relationships between them. After this step, the NSTB proceeds with the image windows shifted in even-numbered blocks, same as in the Swin Transformer \cite{liu2022swin}.}

Our approach differs from that of \cite{10222853}, where the SwinTransformer utilizes implicit window-based self-attention (WSA) to process image patches. This method constrains the receptive field, as it limits the model's ability to capture long-range dependencies beyond the fixed window size. Specifically, the attention is confined within each window, preventing the network from effectively incorporating global context. In contrast, our N-gram refinement technique allows for a more flexible windowing strategy, which enables the model to capture finer details and broader context within the same window. By refining local windows with N-grams, our design expands the effective receptive field, enhancing the model's ability to capture both local and global features. This results in an output image that retains more detailed and comprehensive information, ultimately improving the quality of the image representation.

\vspace{-0.2cm}
\subsection{\textbf{ROI-Weighted Rate Optimization}}
\vspace{-0.2cm}


To balance compression quality and bit rate, we \textcolor{black}{adopt the loss function from \cite{10222853}, which combines distortion in key regions with bit rate control. This loss function includes two terms - one that measures the distortion between the original and compressed images, and another that accounts for the bit rate. The distortion is weighted by the ROI mask, so that important areas in the image are prioritized during compression. The bit rate term ensures that the encoded image remains efficient in terms of size. A trade-off between these two objectives is controlled by weighting factors, where the distortion weight is adaptively adjusted based on the QIndex map. Specifically, this weight depends on a rate parameter that varies according to the maximum and minimum values within the QIndex map, allowing the network to dynamically balance quality and compression.}

\section{Experiments and Results}
\label{sec:experiments}

\subsection{Training and Evaluation}

\par{\textbf{Dataset:}} For training, we use the Flicker 2W dataset, as in \cite{mudgal2024enhancing}, which contains 20,745 high-quality general images, alongside the COCO 2017 \cite{lin2015microsoftcococommonobjects} dataset for ROI-specific training. We randomly select approximately 200 images for validation, while the rest are used for training. The images are cropped into 256 × 256 patches for input. We then train our network on these patches using the CompressAI PyTorch library \cite{bégaint2020compressaipytorchlibraryevaluation}. Note that we exclude images with a height or width smaller than 256 pixels for simplicity. For evaluation, we use the widely recognized Kodak image dataset \cite{kodak}, which contains 24 uncompressed images with a resolution of 768 × 512.

\begin{figure*}[!t]
\vspace{-0.8cm}
\setkeys{Gin}{width=.5\linewidth} 
\begin{minipage}[t]{1\columnwidth}
  \makebox[\linewidth][c]{%
     
    \subfloat[]{\label{fig:kodak_rd_mse}\includegraphics[trim=1cm 0cm 1cm 0cm, clip]{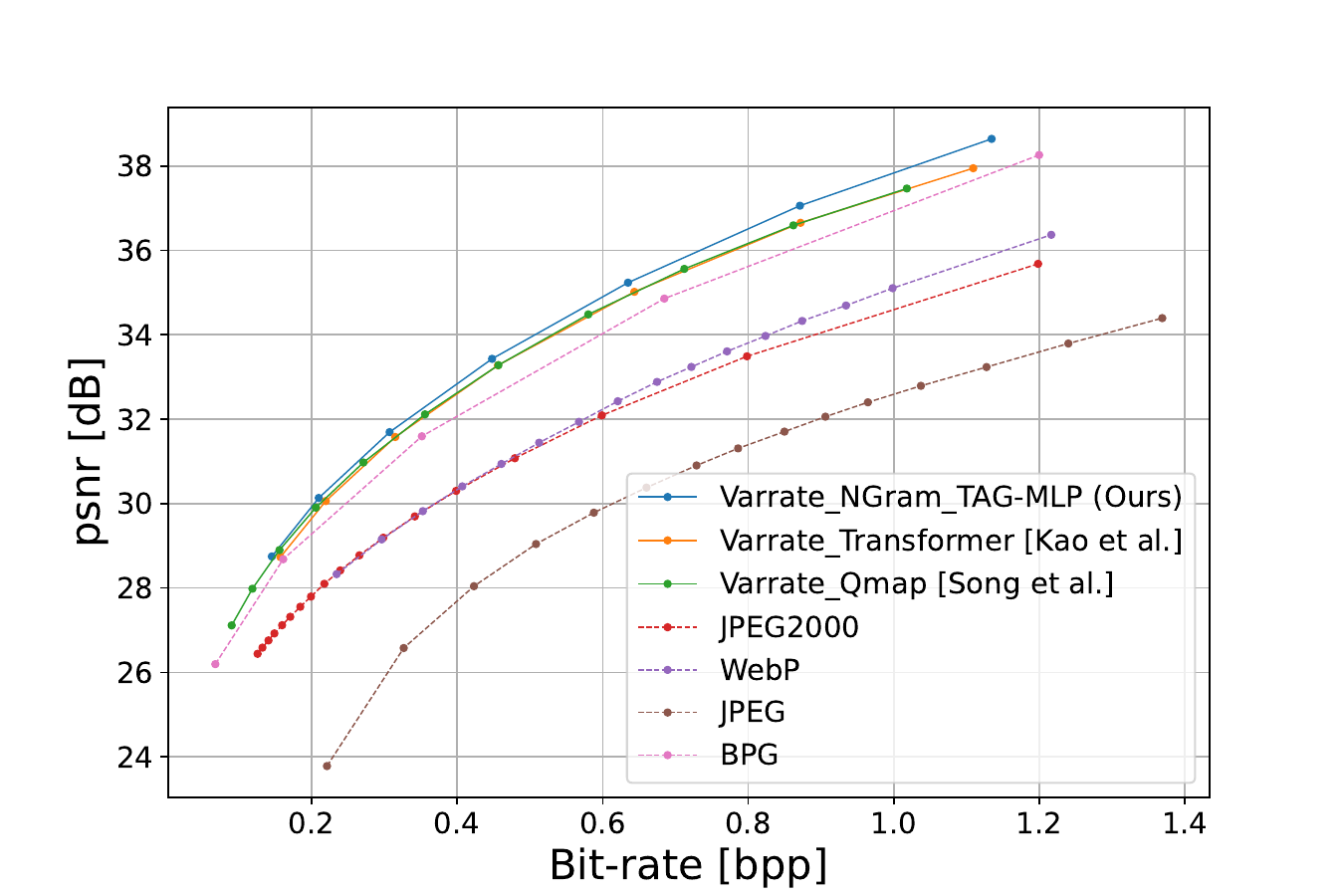}}%
    \hspace{0.5cm}
    \subfloat[]{\label{fig:roi_psnr}\includegraphics[trim = 1cm 0cm 1cm 0cm]{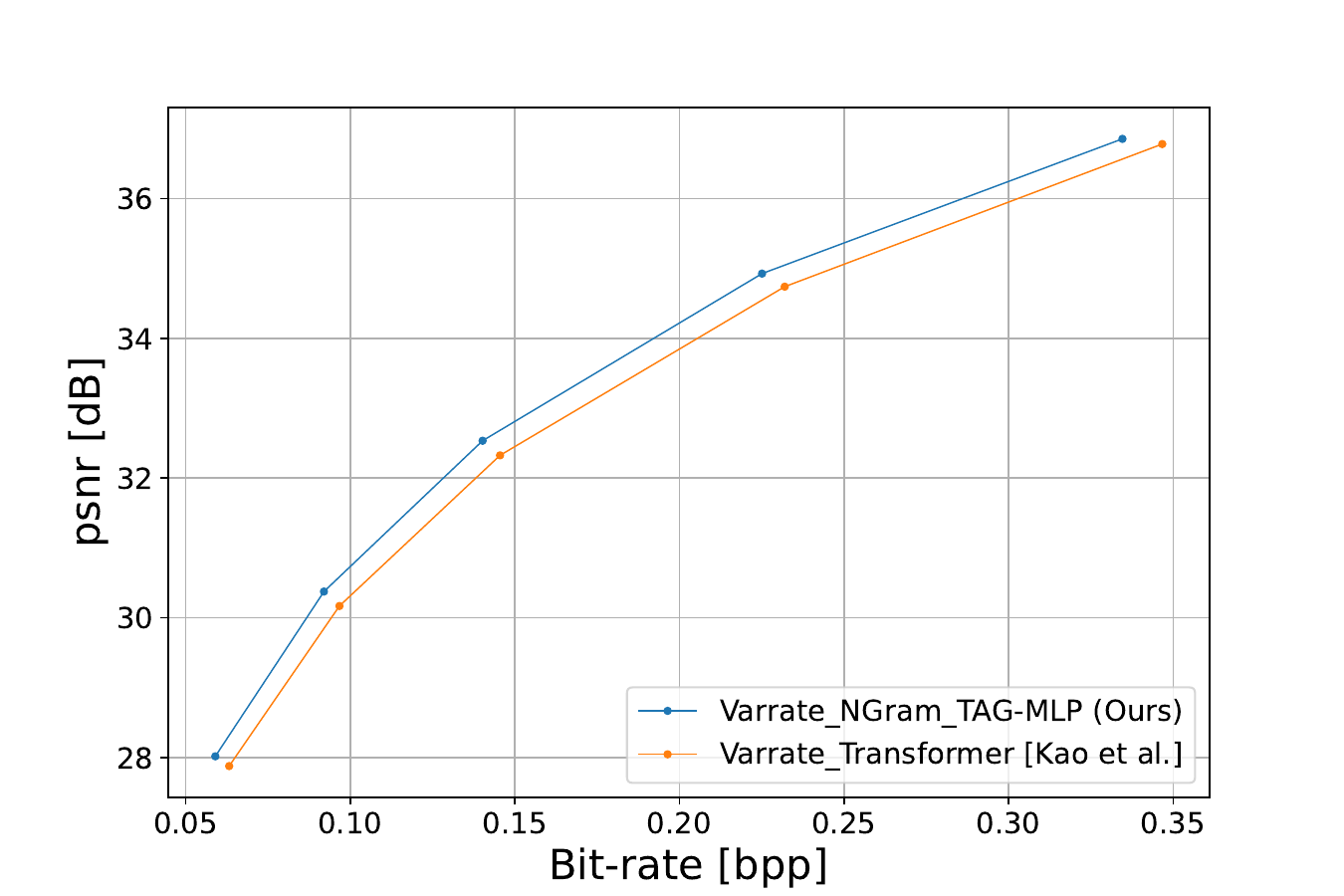}}%
  }
  \makebox[\linewidth][c]{%
    \subfloat[]{\label{fig:nroi_psnr}\includegraphics[trim = 1cm 0cm 1cm 0cm]{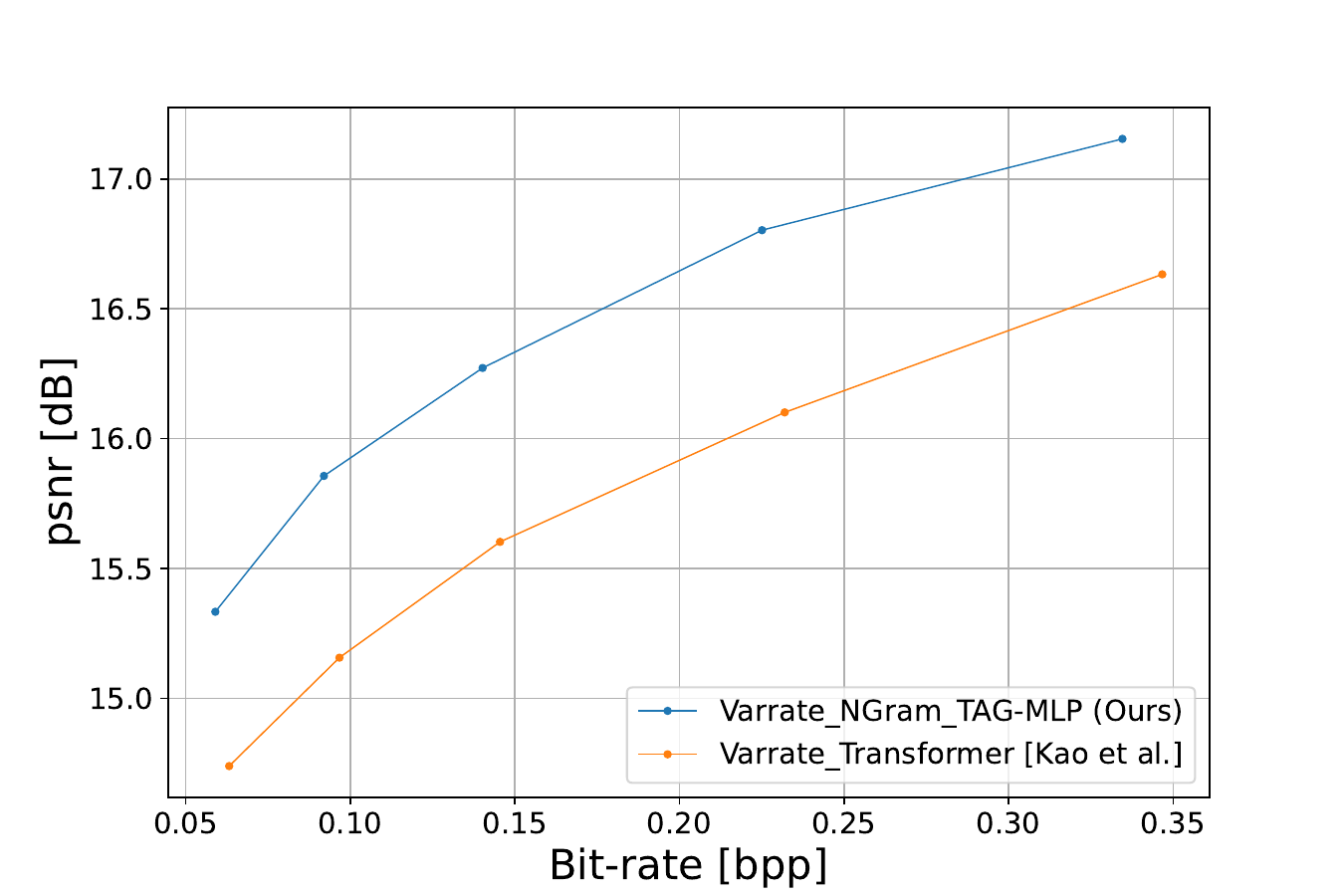}}%
    \hspace{0.5cm}
    \subfloat[]{\label{fig:weighted_psnr}\includegraphics[trim = 1cm 0cm 1cm 0cm]{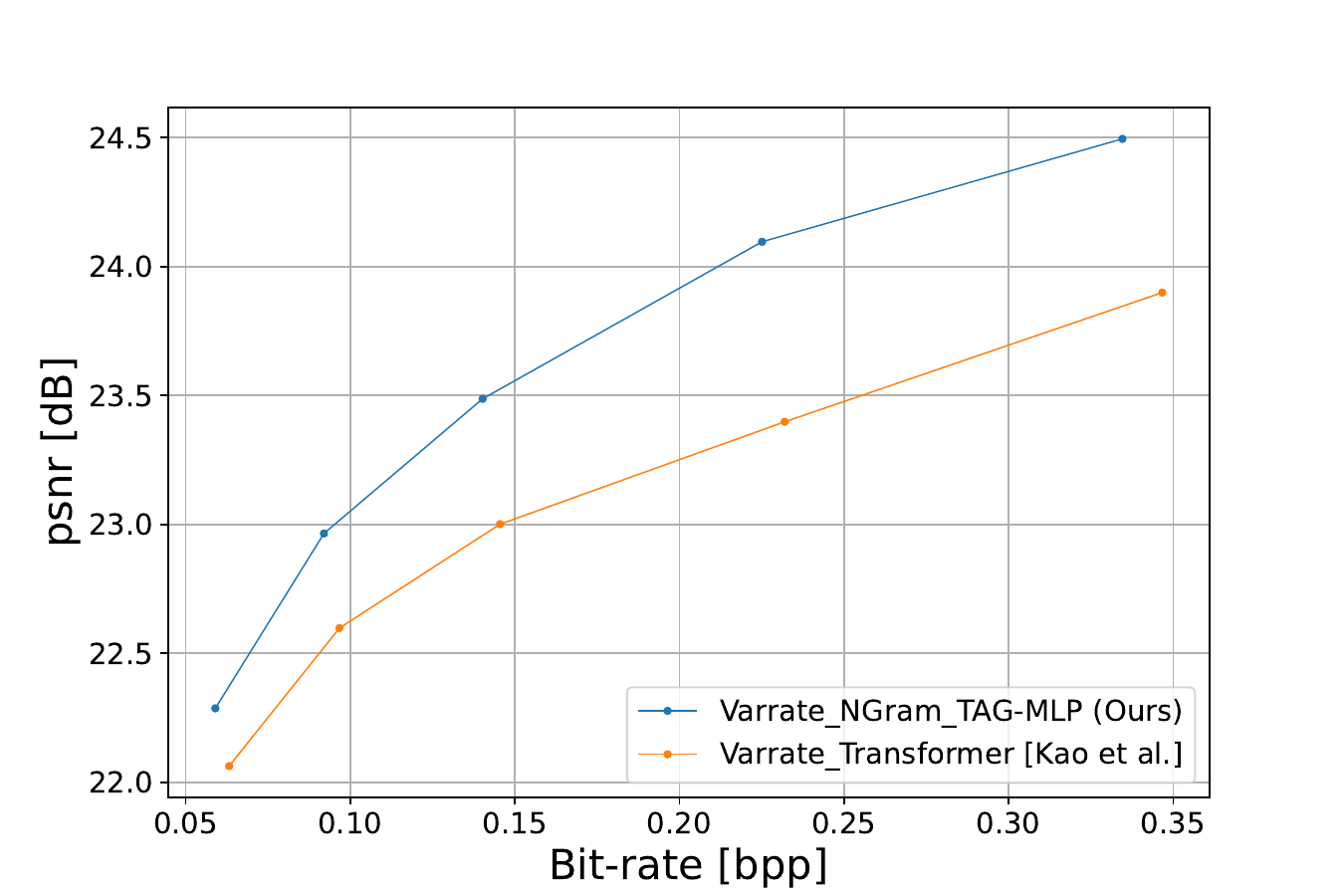}}%
  }
  \caption{RD-performance: (a) Variable-rate coding without ROI on Kodak. (b) Variable-rate coding with ROI on COCO dataset showing the comparison of baseline method \cite{10222853} with our approach. (c) Variable-rate coding with NROI on COCO dataset. (d) Variable-rate coding with ROI approach on full image of COCO dataset.}
  \label{fig:our_results}
\end{minipage}
\end{figure*}
\noindent {\textbf{Implementation:}} All experiments are conducted on a single Nvidia A40 GPU using the Adam optimizer. Following the training scheme from \cite{10222853}, we \textcolor{black}{first} train the model for 400 epochs with the highest $\lambda$ value. \textcolor{black}{Then,} we train for variable-rate coding by sampling $\lambda$ uniformly between $\lambda_{min}$ = 0.0018 and $\lambda_{max}$ = 0.0932 over 350 epochs, using a uniform ROI mask. Finally, we fine-tune for spatial quality control with random ROI masks for 100 epochs. \textcolor{black}{We evaluate the model in two settings: without ROI on the Kodak dataset~\cite{kodak}, and with ROI on the COCO 2017 validation set~\cite{lin2015microsoftcococommonobjects}.} Image quality is measured using weighted PSNR. \textcolor{black}{The corresponding mean squared error (MSE) is computed separately for the ROI and non-ROI (NROI) regions, then combined using a weighted average based on the importance of each region.}

\subsection{Rate-distortion Performance}

\vspace{-0.2cm}

We benchmark our method against state-of-the-art variable-rate image compression models by Kao et al. \cite{10222853}, Song et al. \cite{song2021variableratedeepimagecompression}, and traditional codecs like JPEG \cite{Wallace1991TheJS}, JPEG2000 \cite{Taubman2013JPEG2000I}, WebP \cite{webp}, and BPG \cite{bpg}. We obtain rate-distortion data points for the learned methods from published papers and official GitHub repositories, while results for the traditional methods are from CompressAI’s \cite{bégaint2020compressaipytorchlibraryevaluation} reported benchmarks. We evaluate using PSNR for image distortion and bits per pixel (bpp) for rate, generating RD curves to compare coding efficiency. 

Fig. \ref{fig:kodak_rd_mse} compares state-of-the-art learned methods \cite{10222853,song2021variableratedeepimagecompression} for variable-rate
compression without ROI. Our method, incorporating N-Gram context and TAG-MLP, outperforms them, achieving up to a 0.70 dB PSNR improvement at the highest QIndex on the Kodak dataset \cite{kodak}. Figs. \ref{fig:roi_psnr}, \ref{fig:nroi_psnr}, and \ref{fig:weighted_psnr} show comparisons with the baseline \cite{10222853} in terms of weighted PSNR for ROI, NROI, and full image. Our method consistently outperforms the baseline across all regions, particularly in ROI segments, where the N-Gram context enhances feature interaction and detail preservation, while also improving NROI and overall compression quality.

\begin{figure}
\includegraphics[width=0.9\textwidth, trim = 0cm 4cm 5.5cm 3cm]{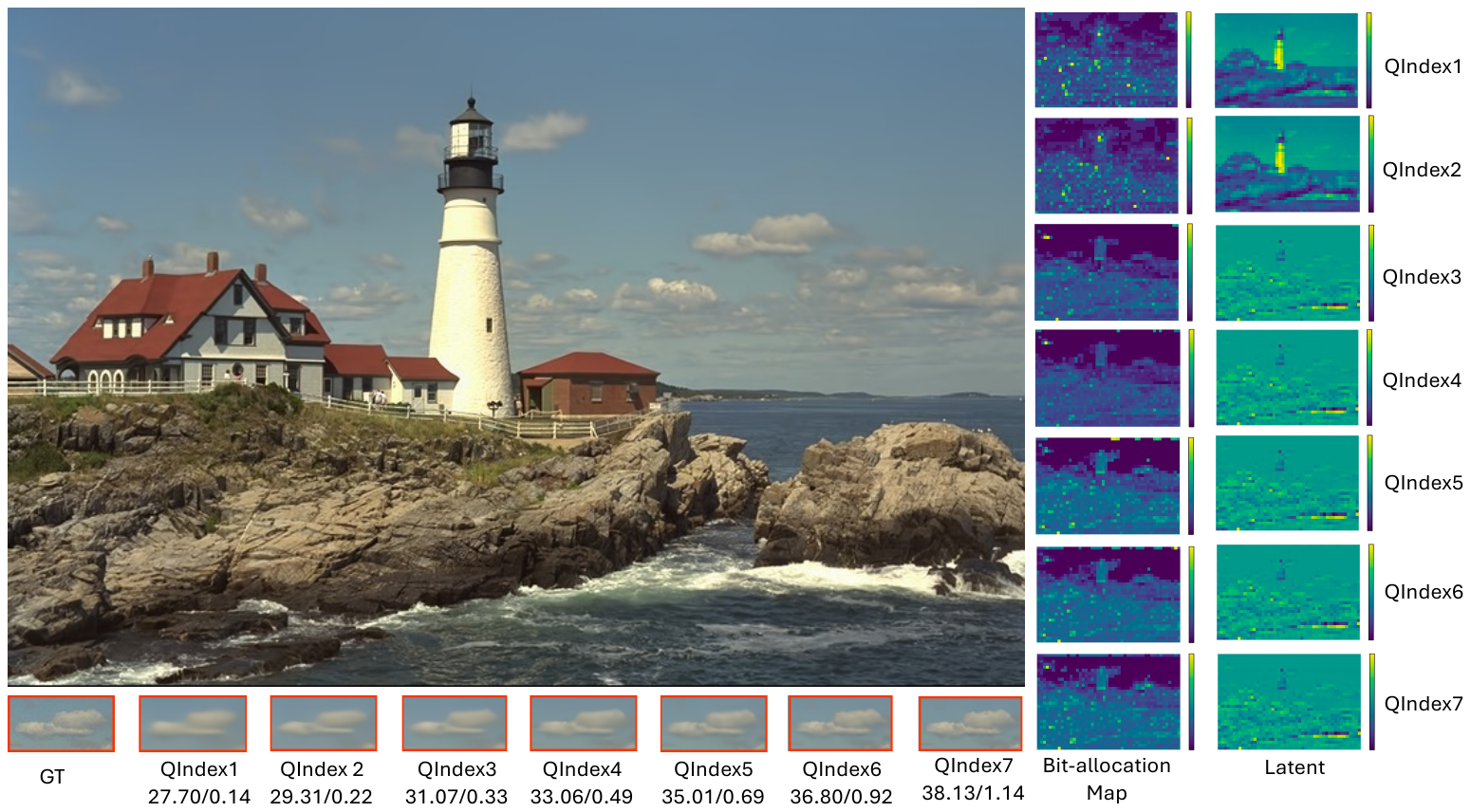}
\caption{\protect\label{fig:kodim01_1}Visualization of our method across different QIndexs and the bit-allocation map for the channel with maximal entropy. The results demonstrate that our approach allocates more bits to high-contrast regions, enhancing their quality, while assigning fewer bits to low-contrast areas, such as the sky and clouds. \textcolor{mycolor}{Corresponding QIndexs, PSNR↑/bpp↓ are mentioned below each image.}}

\end{figure}
\vspace{-0.5cm}
\subsection{Visual Quality} 
\begin{figure}
\centering
\includegraphics[width=0.8\textwidth, trim = 0cm 4.5cm 4.5cm 2.6cm]{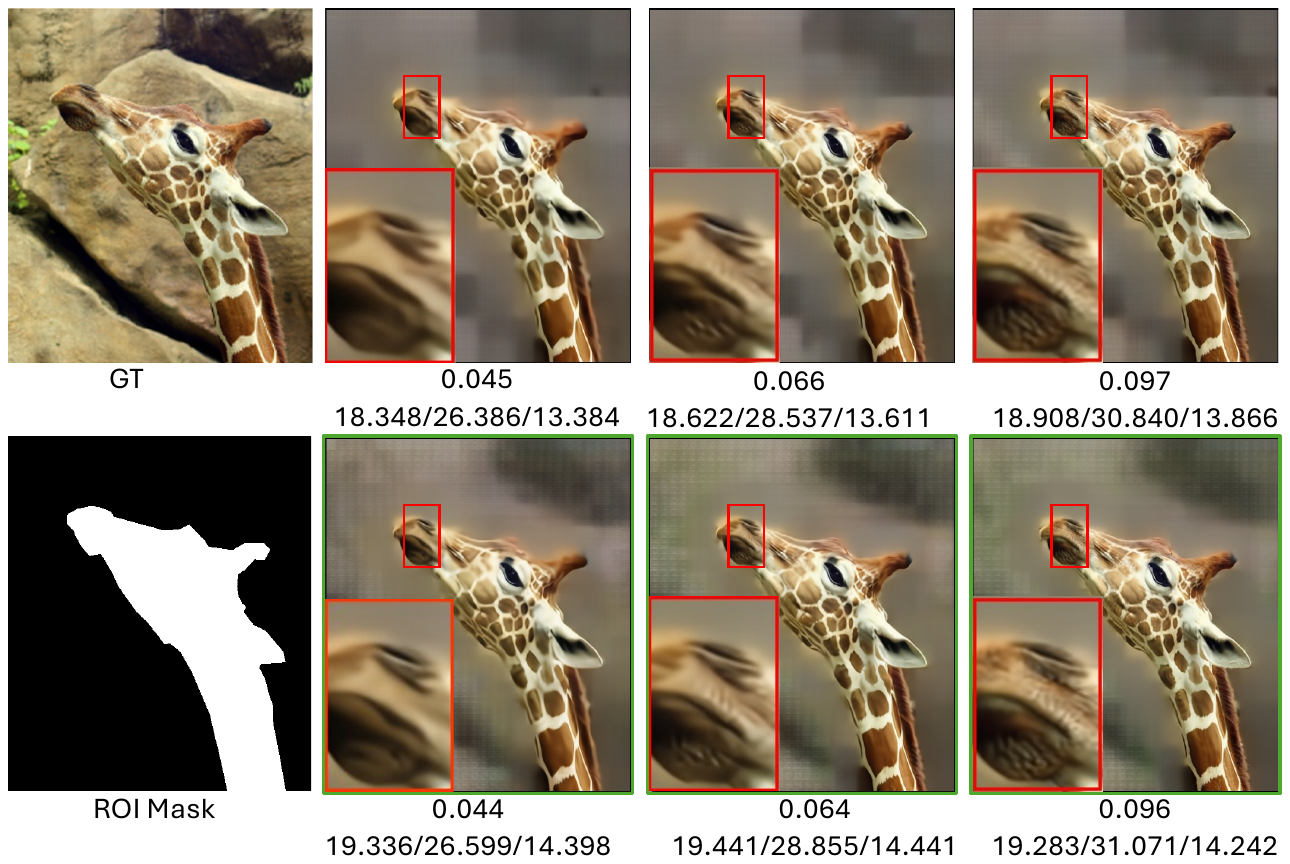}
\caption{\label{fig:roi_comparison}Quality comparison of baseline (top row) \cite{10222853} with our method (bottom row) for ROI segments. Subtitles show corresponding bpp↓ (top in caption) and PSNR↑ (full image/ROI/NROI).}
\end{figure}

\vspace{-0.2cm}
Fig. \ref{fig:baseline_comparison} shows reconstructed images (kodim24.png) using our method, baseline method \cite{10222853}, and compression standards JEPG and WebP. For JPEG and WebP, we target similar bits per pixel (bpp) levels as the learned method.
Our approach retains more details with comparable bpp, resulting in significantly higher PSNR. \textcolor{mycolor}{Fig. \ref{fig:roi_comparison} highlights the superiority of our method over the baseline \cite{10222853}, showing higher PSNR in ROI segments.}  Additionally, in Fig. \ref{fig:kodim01_1}, we show results for kodim21 across seven different quality levels. The images with higher bpp approach the quality of the original image. The bit allocation map for the channel with the highest entropy shows that our method allocates more bits to complex regions and fewer bits to simpler ones as the QIndex increases. 
\subsection{Complexity} 
\vspace{-0.2cm}
We compare the latency of the Kao et al. \cite{10222853} model (32.7M parameters) with our model (33.3M parameters) on the Kodak dataset. \textcolor{black}{Both models were tested on the same system: an NVIDIA A40 GPU with 48GB of memory and an AMD EPYC 7502P 32-Core Processor.} Despite having more parameters, our model achieves a lower latency of 10.9 seconds, compared to 11.12 seconds for baseline model. These results support our hypothesis that the proposed architecture improves processing efficiency. As our design builds on~\cite{10222853}, it inherits swin local attention mechanism, which avoids the computational overhead of global attention. The N-Gram context mechanism used by our method extends the receptive field beyond the baseline, improving RD performance through broader contextual access. Our method achieves improved efficiency through our design aspects of architectural integration. Specifically, NSTB combine scaled-cosine self-attention with N×N average pooling, which lowers computational cost. These design choices contribute to a more efficient pipeline, resulting in slightly faster inference while maintaining high reconstruction quality. However, we acknowledge that further profiling including FLOPs, memory bandwidth usage, and per-layer timing is needed to quantify the precise computational trade-offs. Future work will include a more detailed analysis across varied and larger datasets and hardware setups to validate the generality of these efficiency gains.



\vspace{-0.3cm}

\subsection{Ablation Study} 
\vspace{-0.2cm}

\begin{figure}
\centering
\includegraphics[width=0.8\textwidth, trim = 0cm 0cm -1cm 0cm]{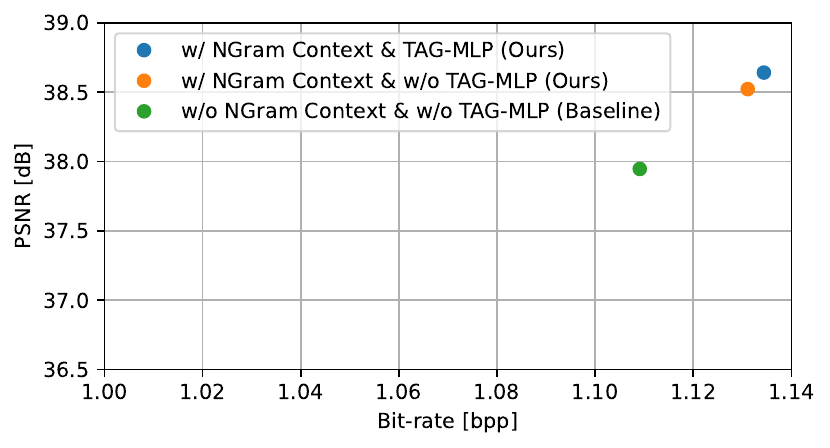}
\vspace{-0.2cm}
\caption{\label{fig:varrate_psnr_comparison_ablation_study}Ablation study of N-gram context and TAG-MLP. We present the results for QIndex 7 with MSE optimization. Our findings show a significant improvement in rate-distortion (RD) performance when using the N-gram context compared to the baseline method \cite{10222853}, with a further slight enhancement when TAG-MLP is incorporated.}
\end{figure}

\textcolor{black}{To better understand the contribution of each component in our architecture, we performed an ablation study by incrementally adding key modules to a baseline model that excludes both N-Gram context partitioning and the TAG-MLP. Starting with the baseline, we observed relatively poor RD performance due to the model’s limited ability to capture contextual dependencies. When we introduced the N-Gram context partitioning mechanism, there was a substantial improvement, as shown in Fig. \ref{fig:varrate_psnr_comparison_ablation_study}. This result highlights the importance of local contextual modeling, the N-Gram mechanism enables the model to better encode short-range dependencies, which are critical for capturing structure in sequential data. Finally, incorporating the TAG-MLP component provided a further, albeit smaller, improvement. We attribute this to TAG-MLP’s ability to better model higher-order interactions across token groups by leveraging their aggregated features more effectively. The combination of both components leads to the best performance, suggesting that N-Gram context captures fine-grained local structure, while TAG-MLP adds complementary capacity to model more abstract or long-range interactions. Together, they form a synergistic architecture that enhances representational power for rate-distortion optimization.}

\vspace{-0.5cm}

\section{Conclusion}
\label{sec:conclusion}
\vspace{-0.2cm}

This paper introduces the novel application of N-Gram context to image compression, enhancing the Swin Transformer with a Sliding-WSA mechanism to address the small receptive field. The integration of N-Gram interactions improves the model's ability to capture long-range dependencies and spatial relationships, leading to better image feature representation and compression. Extensive experiments demonstrate that our approach significantly improves RD-performance, outperforming state-of-the-art methods in both variable-rate and ROI compression. This method enables efficient bit-rate control and adaptive compression for different image regions, making it highly flexible for real-world applications. In future, we see potential for N-Gram context in other tasks like video compression. We set N=2 based on \cite{choi2023ngramswintransformersefficient}, but future work will explore the effect of varying N values on RD performance and optimize the model for larger datasets.

%
%
%
%

\end{document}